\documentclass[twocolumn,superscriptaddress,floatfix,preprintnumbers,prl]{revtex4}

\usepackage{ORI_Group_style}

\begin{document}
  
\title{Ultrashort Pulses for Far-Field Nanoscopy}

\author{Patrick Maurer}
\affiliation{Institute for Quantum Optics and Quantum Information of the
Austrian Academy of Sciences, A-6020 Innsbruck, Austria.}
\affiliation{Institute for Theoretical Physics, University of Innsbruck, A-6020 Innsbruck, Austria.}
\author{J. Ignacio Cirac}
\affiliation{Max-Planck-Institut f\"ur Quantenoptik,
Hans-Kopfermann-Strasse 1, D-85748, Garching, Germany}
\author{Oriol Romero-Isart}  
\email{oriol.romero-isart@uibk.ac.at}
\affiliation{Institute for Quantum Optics and Quantum Information of the
Austrian Academy of Sciences, A-6020 Innsbruck, Austria.}
\affiliation{Institute for Theoretical Physics, University of Innsbruck, A-6020 Innsbruck, Austria.}

\begin{abstract}
We show that ultrashort pulses can be focused, in a particular instant, to a spot size given by the wavelength associated with its spectral width. For attosecond pulses this spot size is within the nanometer scale. Then we show that a two-level system can be left excited after interacting with an ultrashort pulse whose spectral width is larger than the transition frequency, and that the excitation probability depends not on the field amplitude but on the field intensity. The latter makes the excitation profile have the same spot size as the ultrashort pulse. This unusual phenomenon is caused by quantum electrodynamics in the ultrafast light-matter interaction regime since the usually neglected counterrotating terms describing the interaction with the free electromagnetic modes are crucial for making the excitation probability nonzero and depend on the field intensity. These results suggest that a train of coherent attosecond pulses could be used to excite fluorescent markers with nanoscale resolution. The detection of the light emitted after fluorescence—or any other method used to detect the excitation—could then lead to a new scheme for far-field light nanoscopy.
\end{abstract}
\maketitle

Microscopy aims at imprinting features in a sample to discern details within a region as small as possible. While impinging electrons to a sample provides resolution in the nanoscale—namely, nanoscopy—using light is preferable in life sciences due to it being less invasive and to permit tagging parts of the sample with fluorescent markers~\cite{Hell2007}. However, light microscopy has to circumvent the Abbe diffraction limit, which prevents focusing monochromatic light in the far field beyond a spot size of half its wavelength~\cite{Novotny,Born,Kubitscheck,Hell2007}. This can be circumvented (i) using near fields, which are not diffraction limited and can thus be used to achieve higher resolution~\cite{Vigoureux,Trautman,Sanchez,Hillenbrand,Taminiau}. Remarkably, for far-field fluorescent nanoscopy, the Abbe limit can be circumvented by (ii) manipulating bright and dark states of the fluores- cent markers while still using quasimonochromatic diffraction-limited light~\cite{Hell1994,Hell1995,Betzig,Moerner,Rust,Hofmann}. 

Here, we explore an alternative approach for light nanoscopy based on using coherent broadband light, that is, ultrashort pulses~\cite{Chang}. Such pulses have spectral widths comparable—or even larger—in the attosecond regime than optical frequencies~\cite{Krausz,Paul,Kienberger,Corkum,Hassan}. Ultrashort pulses have been used in the context of subwavelength control of nanoplasmonic fields~\cite{Stockman1,Aeschlimann,Stockman2}. In this Letter we address the following questions: What is the minimal spot size of a focused ultrashort pulse? Can an optical transition be excited after the interaction with an ultrashort pulse? Does the excitation profile have the same spot size as the light intensity? The positive answer to these questions allows us to suggest a scheme for far-field light nanoscopy that is based on neither (i) nor (ii).

\begin{figure*}[t]
\centering
\includegraphics[width=  2 \columnwidth]{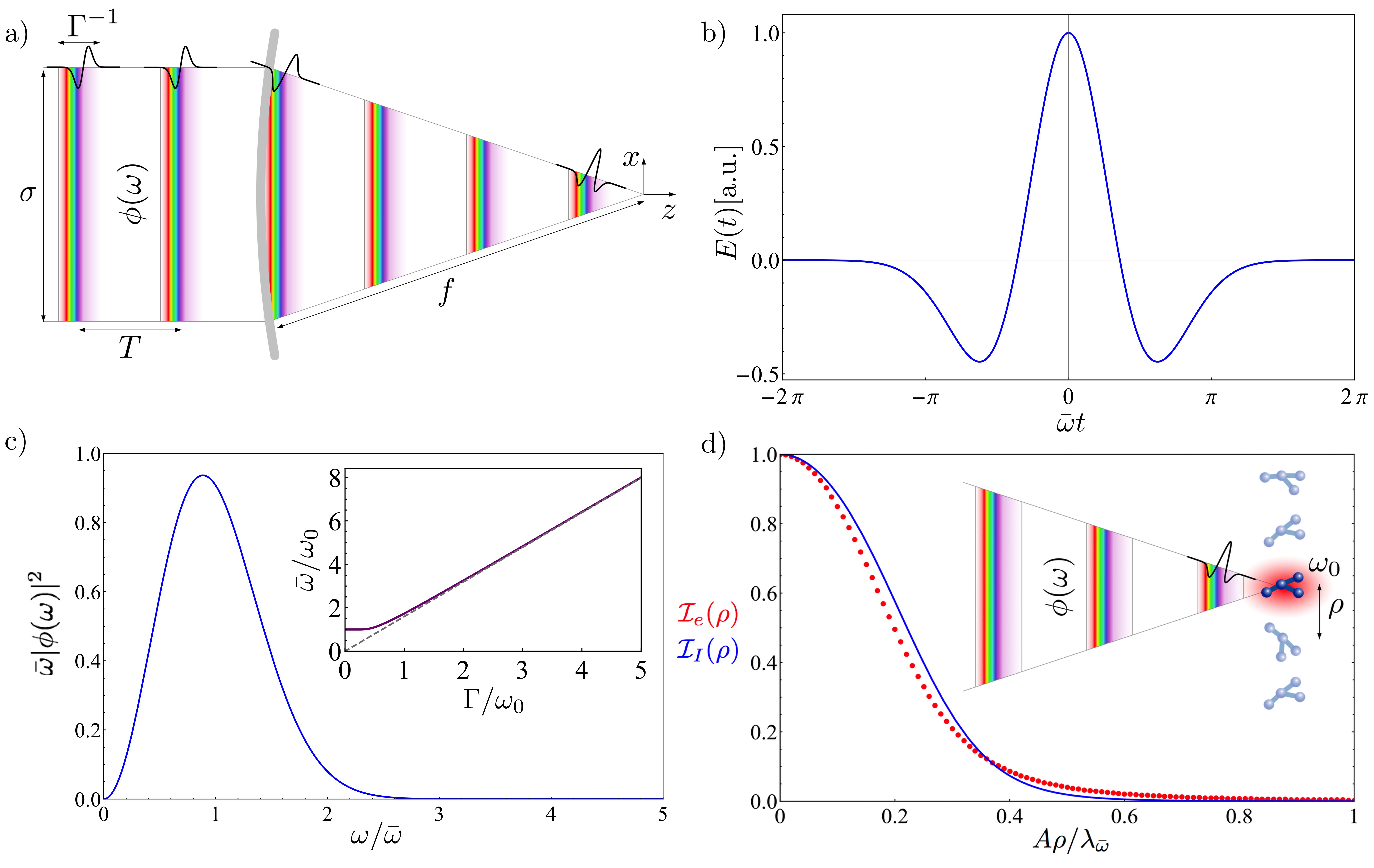}
\caption{ (a) Illustration of the focusing of pulses of duration $\Gamma^{-1}$, spectrum $\phi(\w)$, and waist $\sigma$, which are separated in time by $T$, in a focusing scheme mathematically modeled by a reference sphere of radius $f$. (b) The amplitude of the electric field at the focal point in arbitrary units is plotted as a function of $\bar{\omega} t$ for an ultrashort pulse where $\Gamma\gg\omega_0$.  (c) The spectrum $|\phi(\w)|^2$ in units of $1/\bar{\omega}$ is plotted as a function of $\omega/\bar{\omega}$ in the case of an ultrashort pulse, where $\Gamma\gg\omega_0$. We consider a Gaussian spectrum given by $\phi(\w) = \im \mathcal{N}[l(\w) -l(-\w)]$, with $ l(\w) = \exp[-(\w+\w_0)^2/(4 \Gamma^2)]$ and $\mathcal{N}$, such that $\int_{0}^\infty \text{d} \w |\phi(\w)|^2 = 1$. The inset shows $\bar \w$ in units of $\omega_0$ as a function of $\Gamma/\omega_0$ with the dashed gray line showing the asymptotic value $\bar \w = \Gamma (8/\pi)^{1/2}$. (d) The intensity resolution function $\mathcal{I}_I$ and the excitation resolution function $\mathcal{I}_e$ is plotted as a function of $A \rho/\lambda_{\bar{\omega}}$, where $\lambda_{\bar{\omega}} = 2 \pi c/\bar{\omega}$ and $A=\sigma/f\ll1$ is the numerical aperture.}
\label{Fig1}
\end{figure*}

Let us start by discussing the intensity profile of a focused ultrashort pulse. We consider a nonchirped \cite{Chang} pulse propagating along the $z$ axis and linearly polarized along the $x$ axis. We mathematically model the focusing scheme as a nondispersive reference sphere of radius $f$; see \figref{Fig1}(a). Under the paraxial approximation~\cite{Novotny}, the incident field distribution in the spectral representation on the reference sphere is given by $\EE_\text{in}(\rho,\w) = [2 \mathcal{U}/(\epsilon_0 \sigma^2 c)]^{1/2} \phi(\w) \exp[- \rho^2/\sigma^2]\textbf{e}_x $, where $\sigma$ is the waist of the pulse, $\rho \equiv (x^2+y^2)^{1/2}$ the radial distance, $\mathcal{U} \equiv \epsilon_0 c  \int_\mathds{R} \text{d} \w \int_0^\infty \text{d} \rho \rho \abs{\EE_\text{in}(\rho,\w)}^2/2 $ the energy of the pulse, $\epsilon_0$  the vacuum permittivity, and $c$ the speed of light in vacuum. The spectral function $\phi(\w)$ of the pulse fulfills $\phi(\w) = \phi^\star(-\w)$ and is normalized as $\int_{0}^\infty \text{d} \w \abs{\phi(\w)}^2 = 1$ [see \figref{Fig1}(a)]. We define the mean frequency of the spectral function as $\bar{\omega} \equiv \int_0^\infty \text{d}\omega\,\omega\,|\phi(\omega)|^2$ and the corresponding mean wavelength $\lambda_{\bar{\omega}} \equiv 2\pi c/\bar{\omega}$. The field distribution near the focal point can be obtained by first using  geometric optics to obtain the deflection of the field at the reference sphere~\cite{Novotny}, and then by using Fourier optics to obtain the far field near the focus using the stationary phase approximation~\cite{Mandel,Novotny}. The latter is valid provided the focal point is at the far field for all the relevant frequencies of the pulse, namely $f \gg \lambda_{\bar \w}$. One can then obtain
\be \label{eq:FocusedE}
\EE(\rr,\w) \approx \im  e^{\im \w(f+z)/c} \sqrt{\frac{2\mathcal{U}}{\epsilon_0 c}} \phi(\w) \frac{J_1(  A \w \rho/ c)}{\rho} \uex.
\ee
Note that the radial dependence is given by the Airy disk function $J_1(A\omega\rho)/\rho$, where
$J_1(x)$ is the first order Bessel function of the first kind and  $A \equiv \sigma/f \ll 1$ is the numerical aperture. The Airy disk manifests diffraction, namely the loss of spatial resolution due to the non propagation of evanescent fields. A typical temporal dependence of the electric field in the focal point is plotted in \figref{Fig1}(b).
At the particular time $t=f/c$, which we call the rephasing time, the intensity in the focal plane reads $I(\rho)\propto|\int\text{d}\omega\phi(\omega)J_1(A\omega\rho)/\rho|^2$. The intensity is given by a coherent superposition of Airy disks weighted by the spectral function $\phi(\omega)$. Considering a Gaussian spectral function with a spectral width $\Gamma$ and a carrier-frequency $\omega_0$ [see \figref{Fig1}(c)], the intensity resolution $\mathcal{I}_I(\rho) \equiv 2 I(\rho)/[I(0)+I(\rho)]$ is plotted as a function of $A\rho/\lambda_{\bar{\omega}}$ in \figref{Fig1}(d) as a solid blue line. Note that the spot size $\Delta\rho$, defined by $\mathcal{I}_I(\Delta\rho)\approx 0.5$, is given by
$\Delta\rho \approx 0.2 \lambda_{\bar{\omega}}/A$. As shown in the inset of \figref{Fig1}(c), in the quasimonochromatic case  $\Gamma/\omega_0\ll 1$, one recovers the Abbe limit for monochromatic light, namely $\bar{\omega}\approx\omega_0$. However, for an ultrashort pulse, where $\Gamma/\omega_0\gg 1$, one has $\bar{\omega}\approx\Gamma$. Thus the intensity of an ultrashort pulse containing a broad range of frequencies can be focused,  at the rephasing time, to a spot size given by the wavelength associated with its spectral width.

Let us now address the possibility of using such focused coherent broadband light for far-field nanoscopy. In the following we suggest that this might indeed be possible due to unusual quantum electrodynamical phenomena present in the ultrafast light-matter interaction regime. In particular we discuss the interaction of a two-level system with a coherent train of ultrashort pulses in the ultrafast light-matter interaction regime; specifically, the temporal pulse width is much shorter than the transition period of the two-level system. We show that the counterrotating terms describing the interaction of the two-level system with the free electromagnetic modes induce transitions in which the two-level system is excited and a photon is emitted. Moreover, the probability of leaving the two-level system excited depends on the field intensity. These are surprising and potentially useful results since they could be used to imprint a visible feature with high resolution in a fluorescent marker. In practice, however, the particular multilevel structure of a given fluorescent marker needs to be taken into account and the result will depend very much on its specific electronic structure. We remark that the interaction of a two-level system with a train of ultrashort pulses and resonance fluorescence of an arbitrarily driven two-level system  has already been studied; see, for instance, Refs. \cite{Moreno,Wilkens}. However, these studies focus on the excitation process {\em during} the interaction with the pulse, which can be discussed in the context of optical Bloch equations where the interaction with the free electromagnetic modes is not included. Here we are interested in the state of the two-level system {\em after} the interaction with ultrashort pulses, and the interaction  with the free electromagnetic modes not only is required but also plays a crucial role.

Let us first calculate the  probability of exciting a two-level system $\cpare{ \ket{g},\ket{e} }$ situated in the focal region after interacting with a train of $N$ coherent ultrashort pulses, each with spectral width $\Gamma$ and energy $\mathcal{U}$ and separated in time by $T \gg 1/\Gamma$; see \figref{Fig1}(a). We consider an optical transition of frequency $\omega_0$, a free space spontaneous emission rate $\Gamma_0$, and a dephasing rate $\gamma=\Gamma_0/2+\gamma_c$, with $\gamma_c$ being the inhomogeneous broadening of the optical transition. We consider $\gamma N T \ll 1$, such that the dynamics can be approximated to be purely unitary. 
In the displaced classical frame where all of the electromagnetic field modes are in vacuum, the light-matter interaction can be described by the Hamiltonian $\Hop_\text{tot}  = \Hop_0 + \hat V(t) + \Hop_\text{int}$. Here, $\Hop_0$ is the free part of the Hamiltonian  given by
$\Hop_0 =\hbar \w_0 \ketbra{e}{e} + \int_{\mathds{R}^3} \text{d}\kk \sum_{\epsilon} \hbar \w_k \adop_\epsilon (\kk) \aop_\epsilon (\kk)$, where $\aop_\epsilon(\kk)$ [$\aop^\dagger_\epsilon(\kk)$] is the annihilation (creation) operator of the free electromagnetic field mode of wave vector $\kk$ and polarization $\boldsymbol{\epsilon} \perp \kk$, which fulfills $\coms{\aop_\epsilon(\kk)}{\aop^\dagger_{\epsilon'}(\kk')} = \delta(\kk - \kk') \delta_{\epsilon \epsilon'}$. The frequency of the free modes is $\w_k = ck$, where  $k=|\kk|$. The term $\hat V(t)$ is the time dependent part of the Hamiltonian that describes the interaction of the ultrashort pulses with the two-level system and is given by
$\hat V(\rr,t)=\hbar\Omega(\rr,t) \sx$, where $\sx=\spl + \smi$, with $\spl =( \smi)^\dagger = \ket{e} \bra{g}$. We assume that $\cpare{\ket{g},\ket{e}}$ are levels with a zero magnetic quantum number, so that only the component along the x axis of the dipole operator $\hat{\dd}$ is real and nonzero~\cite{Cohen}. Therefore, the Rabi frequency is given by $\Omega\rt = -\sum_{s=0}^{N-1} d_{eg}E(\rr,t-s T)/\hbar\equiv \sum_{s=0}^{N-1} \Omega_0(\rr,t-s T)$, where $\EE(\rr,t)=E(\rr,t)\boldsymbol{e}_x$ is the electric field of a single pulse and $d_{eg}=\bra{e}\hat{\boldsymbol{d}}\ket{g}$. The term $\Hop_\text{int}$ describes the coupling of the two-level system with the free electromagnetic modes, namely
$
\Hop_\text{int} =\im \hbar  \int_{\mathds{R}^3} \text{d}\kk \sum_{\epsilon}g_{k\epsilon} \spare{\aop_\epsilon (\kk)e^{\im\kk\cdot\rr} - \hc} \sx,
$
where the coupling rate is given by $g_{k\epsilon}\equiv -d_{eg} [\w_k/(2 \epsilon_0 (2 \pi)^3\hbar)]^{1/2}\boldsymbol{\epsilon} \cdot  \boldsymbol{e}_x$. Using the total Hamiltonian $\Hop_\text{tot}$, we calculate below the transition probability to excite the two-level system.

We consider that, at $t=t_0$, before the first pulse has interacted with the two-level system, the initial state is given by $\ket{\psi_i}  = \ket{g} \otimes \ket{0} \equiv \ket{g,0}$, where $\aop_\epsilon (\kk) \ket{0} = 0$ $\forall\,\kk, \boldsymbol{\epsilon}$. We are interested in the transition probability, after the interaction with the train of ultrashort pulses, to the state $
\ket{\psi_f}  =   \ket{e} \otimes \adop_\epsilon(\kk)  \ket{0} \equiv \ket{e,\kk \epsilon}$ namely, the two-level system in the excited state and a single photon in the mode with wave vector  $\kk$ and polarization $ \boldsymbol{\epsilon}$. The natural transition to the state $
\ket{\psi_f}  =   \ket{e,0}$, where no photons have been emitted, will also be considered. In first order perturbation theory with $\Hop_\text{int}$, the transition amplitude $c(\rr,t)$ from state $\ket{\psi_i}$ to state $\ket{\psi_f}$ is given by $c(\rr,t)\approx c_0(\rr,t)+c_{\kk \epsilon}(\rr,t)$, where $c_0(\rr,t) \equiv \bra{e,0}\hat{U}(\rr,t,t_0)\ket{g,0}$ accounts for the transitions in which no photons have been emitted and $
c_{\kk \epsilon}(\rr,t) \equiv (\im\hbar)^{-1} \int_{t_0}^t \text{d} t' \bra{e,\textbf{k}\boldsymbol{\epsilon}} \hat U(\rr,t,t') \Hop_\text{int} \hat U(\rr,t',t_0) \ket{g,0}$ for the transitions in which the two-level system is excited and a free photon is emitted. The unitary time evolution operator $\hat{U}(\rr,t,t_0)$ is given by the Schrödinger equation $\im\hbar\partial_t\hat{U}(\rr,t,t_0)=[\hat{H}_0+\hat{V}(t)]\hat{U}(\rr,t,t_0)$, with the boundary condition $\hat U(\rr,t_0,t_0)=\mathds{1}$. In standard light-matter interaction regimes, one would expect the transition amplitude $c_0$ to dominate over $c_{\kk \epsilon}$. However, this is not the case in the ultrafast interaction regime defined by $\Gamma \gg \w_0$. In this regime one can use sudden perturbation theory~\cite{Galindo} to approximate $\exp[- \Hop_0 t /(\im \hbar)] \hat V(t) \exp[ \Hop_0  t/(\im \hbar)] \approx \hat V(t) $, which then, together with $[\hat V(t),\hat V(t')]=0$, can be readily used to integrate the Schr\"odinger equation for $\hat U(\rr,t,t')$ and to obtain  $
\hat U (\rr,t,t_0) \approx \exp [\Hop_0 t/(\im \hbar)] \exp [\chi(\rr,t,t_0)\sx/\im] \exp[-\Hop_0 t_0/(\im \hbar)]$, where we have defined
$\chi (\rr,t,t_0)\equiv\sum_{s=0}^{N-1}\chi_s(\rr,t,t_0)$, with $\chi_s(\rr,t,t_0)\equiv \int_{t_0}^t
dt' \Omega_0(\rr,t'-sT)$. Using this, one can straightforwardly show that $c_0(\rr,t)|_{t=\infty} = 0$ because it depends on the pulse spectrum at zero frequency $\phi(0)$, which has to be zero since ultrashort pulses do not carry electrostatic fields~\cite{Schultz}. The probability to excite the two-level system is thus given by $p_e(\rr)=\lim_{t\rightarrow\infty}\int_{\mathds{R}^3}\text{d}\kk\sum_{\epsilon}|c_{\kk \epsilon}(\rr,t)|^2$ and is only given by transitions in which a photon is also emitted. Note that, since the transition amplitude in zero order in $\Hop_\text{int}$, namely $c_0$, is zero, the transition amplitudes in second or higher orders in $\Hop_\text{int}$, which would excite the two-level system and emit two or more photons, can be safely neglected.

We remark that the dominant transitions that excite the two-level system by emitting a single photon are enabled by counterrotating terms that would typically be neglected under the rotating-wave approximation with respect to the free electromagnetic modes, which is not valid in the ultrafast interaction regime given by $\Gamma \gg \w_0$. Should one not consider the interaction with the free electromagnetic modes, one would incorrectly conclude that the probability of exciting the two-level system after the interaction with the pulses is zero.  Note that the photon emitted during the excitation process should not be confused with the later spontaneously emitted fluorescent photon. 

In order to get a simplified expression for the excitation probability $p_e(\rr)$, we evaluate the expression in the weak field regime, namely $\eta \equiv \max_t\chi(\rr,t)\ll 1$, where $\chi(\rr,t)\equiv\chi(\rr,t,-\infty)$, and that the temporal distance between pulses $T$ is a multiple of $2\pi/\omega_0$. One then obtains
$
p_{e}(\rr)\approx f(\rr) 2N\Gamma_0/(\pi\omega_0^3),
$
where 
\be \label{eq2}
f(\rr) \equiv \int_{0}^{\infty}\text{d}\omega_k \omega_k^3\left|\int_{\mathds{R}}\text{d}\tau e^{\im \omega_k \tau}\sin(\omega_0\tau)\chi_0^2(\rr,\tau)\right|^2
\ee
Using a Gaussian spectral function [recall \figref{Fig1}(b)], the probability of exciting a two-level system situated at the focus is given by $p_e(\rr=0) \approx 25.3 \, \eta^4 N \Gamma_0/\w_0 $, with $\eta \approx 0.64 A [\mathcal{U} \Gamma_0/(\hbar \w_0 \Gamma)]^{1/2}$. 
After the interaction with each train of pulses, a fluorescence emission from the two-level system could be detected during a time scale $1/\Gamma_0$. Thus, the imaging rate is defined as  
$ R \equiv  p_e/(NT+1/\Gamma_0) $. 
We show below that $R$ is of the order of $10^5$ \text{Hz} for typical experimental parameters with attosecond pulses and an optical transition. 

Note that $p_{e}(\rr)$ depends on the light intensity through its dependence on $\chi_0^2(\rr,t)$ in \eqnref{eq2}. This is a consequence of the contribution of the counterrotating terms describing the interaction with the free modes, with a crucial implication for nanoscopy purposes. Let us define, similarly to the intensity resolution, the excitation resolution $\mathcal{I}_e(\rho)= 2 p_e(\rho)/[p_e(0)+p_e(\rho)]$. The dotted red line in \figref{Fig1}(d) shows the excitation resolution as a function of $A\rho/\lambda_{\omega}$. The excitation resolution has a spot size set by $\Delta\rho\approx 0.4\pi c/(A \bar \w)$, similar to the intensity. Hence, a two-level system with transition $\w_0$ can be excited using ultrashort pulses with $\Gamma \gg \w_0$ within a spatial resolution of $ 0.4\pi c/(A \Gamma) \ll 0.4\pi c/(A  \w_0)$, and thus its emission would discern details well beyond the wavelength associated to $\w_0$. We remark that these results do not qualitatively depend on the particular functional form of the pulse spectrum. Note that such a high resolution limit is achieved in the focal plane but not along the optical axis where the pulses propagate.

In order to get an order of magnitude, consider a red optical transition of $2 \pi c/\w_0 = 719$ nm, a lifetime of $\Gamma_0^{-1}=1.6$ ns~\cite{Atto700}, and an inhomogeneous broadening $\gamma_c = 10 \Gamma_0$, which interacts with a train of $N=453$ (such that $N \gamma T =0.1$) coherent Gaussian pulses of $\Gamma^{-1}=38  \times 10^{-18}$ s ($\Gamma/\w_0=10$), each separated by $T=33.6 \times 10^{-15}$ s ($2\pi T/ \w_0 =14$), and with an energy of $\mathcal{U}=0.7$ nJ (such that $\eta=0.5$). The pulses are focused in a scheme with the  numerical aperture $A=0.1$. This leads to an imaging rate $R=1.1\times 10^5$ Hz and a resolution limit of $\Delta \rho = 90$ nm. A train of attosecond pulses with a stable pulse-to-pulse carrier envelope phase and with one pulse per infrared cycle can be produced using high harmonic generation~\cite{Mauritsson,Lewenstein,Ferray}. Our proposal requires a focusing scheme able to prepare a field similar to \eqnref{eq:FocusedE} near the focal point. This could be achieved, for instance, either using sufficiently broadband mirrors or using dispersive elements with optimally chirped pulses.

Our results indicate that focused ultrashort pulses can be used for far-field nanoscopy by imprinting features in a region of a size given by the wavelength associated with their spectral bandwidth. For the particular case of fluoresence nanoscopy~\cite{Hell2007}, we have shown that a two-level marker with an optical transition can be excited due to the usually neglected counterrotating terms describing the interaction with the free electromagnetic modes that are crucial in the ultrafast light-matter interaction regime. Independently of the nanoscopy purposes, it would be very interesting to experimentally show the discussed excitation process as a manifestation of quantum electrodynamics in the ultrafast interaction regime $\Gamma\gg\omega_0$. This could, in principle, be observed either using attosecond pulses and optical transitions or femtosecond pulses and radio-frequency transitions.
In practice, one should consider the particular multilevel structure of the fluorescent marker; see Ref.\cite{Matveev}. Fast transitions from higher levels could increase the probability of exciting the optical level, and cascade transitions could emit other colors. While this can be  analyzed with some of the theoretical tools given here, the results will strongly depend on the particular electronic structure of the molecule. In any case, it seems plausible that one can find molecules where visible features would still be imprinted with such a high resolution.  We hope that our results will trigger further research into the interplay between the fields of attosecond physics and optical microscopy.

We acknowledge our discussions with L\'aszl\'o Veisz and Romain Quidant. This work is supported by the Austrian Federal Ministry of Science, Research, and Economy (BMWFW).

\end{document}